# Quantum evolution of the disoriented chiral condensates

Y. Kluger[a], F. Cooper, E. Mottola, J.P. Paz [b] and A. Kovner[c]

[a]Nuclear Science Division, Lawrence Berkeley Laboratory,
MS 70A-3307, Berkeley, CA 94720, USA

[b]Theoretical Division, Los Alamos National Laboratory,
Los Alamos, New Mexico 87545 USA

[c]School of Physics and Astronomy, University of Minnesota,
Minneapolis, Minnesota 55455, USA

We study the dynamics of the chiral phase transition expected during the expansion of the quark-gluon plasma produced in a high energy hadron or heavy ion collision, using the $O(4)$ linear sigma model in the mean field approximation. Imposing boost invariant initial conditions at an initial proper time $\tau_0$ and starting from an approximate equilibrium configuration, we investigate the possibility of formation of disoriented chiral condensate during the expansion. In order to create large domains of disoriented chiral condensates low-momentum instabilities have to last for long enough periods of time. Our simulations show no instabilities for an initial thermal configuration. For some of the out-of-equilibrium initial states studied, the fluctuation in the number of particles with low transverse momenta become large at late proper times.

## 1. MODEL AND APPROXIMATIONS

The possibility of producing large correlated regions of quark condensate $<\bar{q}_i q_j>$ pointing along the wrong direction in isospin space was proposed [1] to explain rare events where there is a deficit or excess of neutral pions observed in cosmic ray experiments [2]. The idea that such disoriented chiral condensates (DCC's) can be formed has been the source of several experimental proposals [3] in high energy collisions. It has been recognized that nonequilibrium dynamics can yield regions of DCC [4-6] (see Gavin and Asakawa contributions in these proceedings).

There are clearly two important questions that the theoretical models should aim to answer. The first one is to determine whether during the evolution that follows the collision there are instabilities affecting the fluctuations, in which case there is a chance of the correlations growing. The second question is, assuming the instability exists, can the correlated domains grow large enough so that many pions can be emitted from each domain making the detection of DCC's possible.

To address these questions we employ the $O(4)$ linear sigma model, using boost invariant initial conditions at an initial proper time $\tau_0$, in the mean field approximation. This model seems to have the essential attributes of being simple but realistic enough: it has the correct chiral symmetry properties, it also appropriately describes the low energy phenomenology of pions and dynamics then incorporates a cooling mechanism that may lead to low momentum instabilities. Moreover, the mean field approximation *improves* the classical approximation, because it takes into account quantum fluctuations that can not be neglected when the system cools down, and because it enables us to estimate the number of pions and its fluctuations at late times when the condensate settles down near to its vacuum expectation value.

In this model the fields are organized in an $O(4)$ vector $(\phi_1, \phi_2, \phi_3, \phi_4) = (\sigma, \pi_1, \pi_2, \pi_3)$ and we use the following set of coordinates

$$\tau \equiv (t^2 - z^2)^{1/2}, \qquad \eta \equiv \frac{1}{2} \log(\frac{t+z}{t-z}), \qquad \mathbf{x}_\perp, \tag{1}$$

where $\tau$ is the proper time and $\eta$ is the spatial rapidity. The equations of motion are given by

$$\tau^{-1} \partial_\tau \tau \partial_\tau \bar{\phi}_i(\tau) + \bar{\chi}(\tau) \bar{\phi}_i(\tau) = H \delta_{i1}, \tag{2}$$

$$\bar{\chi}(\tau) = \lambda(-v^2 + \sum_{i=1}^{N}[\bar{\phi}_i^{\ 2}(\tau) + <\varphi^2(\eta, x_\perp, \tau)>]), \tag{3}$$

$$\left(\tau^{-1} \partial_\tau \tau \partial_\tau - \tau^{-2} \partial_\eta^2 - \partial_\perp^2 + \bar{\chi}(\tau)\right) \varphi(\eta, x_\perp, \tau) = 0, \tag{4}$$

where $\chi = \lambda(\sum_{i=1}^{N} \phi_i \phi_i - v^2)$ with $N = 4$. At equal proper times the quantum field $\varphi_i(x, \tau)$ defined here satisfies

$$< \varphi(\eta_x, x_\perp, \tau) \, \varphi(\eta_y, y_\perp, \tau) > = G_c(\eta_x, x_\perp, \eta_y, y_\perp; \tau), \tag{5}$$

where $G_c(x, y) \equiv \frac{1}{N} \sum_{i=1}^{N} < \phi_i(x) \phi_i(y) > - \bar{\phi}_i(x) \bar{\phi}_i(y)$. We expand the field $\varphi$ in an orthonormal basis

$$\varphi(\eta, x_\perp, \tau) \equiv \frac{1}{\tau^{1/2}} \int dk (\exp(ik \cdot \mathbf{x}) f_k(\tau) \, a_k + h.c.), \tag{6}$$

where $k \cdot \mathbf{x} \equiv k_\eta \eta + \vec{k}_\perp \vec{x}_\perp$, $dk \equiv dk_\eta d^2 k_\perp / (2\pi)^3$, and the mode functions $f_k(\tau)$ evolve according to:

$$\ddot{f}_k = -(\frac{k_\eta^2}{\tau^2} + p^2 + \bar{\chi}(\tau) + \frac{1}{4\tau^2}) f_k \equiv -\omega_k^2 f_k; \qquad p \equiv |\vec{k}_\perp|. \tag{7}$$

A dot here denotes the derivative with respect to the proper time $\tau$. The Fourier transform of the expectation value $< (\varphi)^2(x, \tau) >$ can be expressed in terms of the mode functions $f_k$ and of the distribution functions

$$n_k \equiv < a_k^\dagger a_k >, \quad g_k \equiv < a_k a_k >, \quad g_k^* \equiv < a_k^\dagger a_k^\dagger > \tag{8}$$

which entirely characterize the initial state of the quantum field.

In order to solve equations (2), (3) and (7) as an initial value problem, we need to fix at $\tau_0$ the mean values $\bar{\phi}_{ik}$, $\dot{\bar{\phi}}_{ik}$, $n_k$, $g_k$ and $g_k^*$. In the boost invariant case $\bar{\phi}_{ik}(\tau) = \dot{\bar{\phi}}_{ik}(\tau) = 0$ for $k \neq 0$.

The complexity of the processes we study suggests that our initial configuration is not a pure state but a mixed ensemble that should be described by a density matrix. In the mean field approximation the most general density matrix for each mode $k$ is a gaussian with five parameters[1] associated with the above initial conditions and it can be written as

$$\rho(\phi_{ik}, \phi'_{ik}) = \frac{1}{\sqrt{2\pi G_{ik}}} \exp\{i\dot{\bar{\phi}}_{ik}(\phi'^{*}_{ik} - \phi^{*}_{ik}) - \frac{1}{2}(K_{ik} - 4G_{ik}\Sigma^2_{ik})|\phi'_{ik} - \phi_{ik}|^2$$
$$-i\Sigma_{ik}[|\phi_{ik} - \bar{\phi}_{ik}|^2 - |\phi'_{ik} - \bar{\phi}_{ik}|^2] - \frac{1}{8G_{ik}}|\phi_{ik} + \phi'_{ik} - 2\bar{\phi}_{ik}|^2\}, \qquad (9)$$

where

$$G_{ik} = <\phi_{ik}\phi^*_{ik}> - \bar{\phi}_{ik}\bar{\phi}^*_{ik} \quad ; K_{ik} = <\dot{\phi}_{ik}\dot{\phi}^*_{ik}> - \dot{\bar{\phi}}_{ik}\dot{\bar{\phi}}^*_{ik} \qquad (10)$$

$$4\Sigma_{ik}G_{ik} = <\{\phi_{ik}, \dot{\phi}^*_{ik}\}> - \{\bar{\phi}_{ik}, \dot{\bar{\phi}}^*_{ik}\}. \qquad (11)$$

Pure density matrix satisfies the condition $p \equiv K - 4G\Sigma^2 + 1/4G = 0$ (no $\phi_i(k)\phi'_i(k)$ mixed term). This subset corresponds to squeezed-coherent states [7]. For $\Sigma = 0$ and $G_{ik} = 1/\sqrt{2\omega_k}$, the density matrix operator becomes $\hat{\rho}^c = |c><c|$ where $|c>$ is a coherent state. One can make a direct connection between our calculation and earlier work [4], which used classical approximation to the same model. The equilibrium density matrix of a free theory is Gaussian and in the high $T$ limit ($\beta\omega_k << 1$) has the following representation:

$$\rho = \prod_{ik} \int d\bar{\phi}_{ik} d\dot{\bar{\phi}}_{ik} \exp\{-\frac{\beta}{2}(|\dot{\bar{\phi}}_{ik}|^2 + (k^2 + m^2)|\bar{\phi}_{ik}|^2)\}\rho^c_{ik}(\bar{\phi}_{ik}, \dot{\bar{\phi}}_{ik}). \qquad (12)$$

Therefore, solving the mean field equations with an initial thermal homogeneous density matrix is equivalent to solving the semi classical equations for a thermal ensemble of inhomogeneous "classical" initial conditions.

## 2. RESULTS

Fig. 1 shows the mean density of particles $<n(p)> \equiv p \int dk_\eta < a^\dagger_k(t) a_k(t)>/(2\pi)^3 \equiv p \int dk_\eta n_k(t)/(2\pi)^3$ and the fluctuation $\Delta n(p) \equiv p \int dk_\eta < (n_k(t) - <n_k(t)>)^2>/(2\pi)^3$ at the initial proper time $\tau_0 = 1 fm/c$ and at $\tau = 10, 20 fm/c$. The figures on the left correspond to an initial thermal configuration. In this case $\bar{\chi}$ does not become negative during the evolution so no instabilities develop. The density $<n(p)>$ hardly changes during the evolution, as does its fluctuation. The fluctuation at large proper times and at the initial proper time are of the same order of magnitude. Therefore, observing a DCC for an initial thermal ensemble is unlikely. In the right column we generated instabilities by choosing an initial ensemble that is far from equilibrium for which we modified only the initial condition for $\dot{\bar{\sigma}}$ from 0 to $-1$. In this case $\bar{\chi}$ becomes negative at early stages of the evolution, and an exponential growth of the low momenta modes occurs. The density $<n(p)>$ is strongly amplified in the region of small transverse momenta by acquiring energy from the $\bar{\sigma}$ field. The fluctuation of the number density also grows substantially for small transverse momenta, and the observation of DCC is more likely for this "peculiar" ensemble. For similar out-of-equilibrium ensembles we found no instabilities if we start the simulations at proper times greater than $2 fm/c$. This implies that if the expansion rate is too slow or if we start with lower energy densities even for out of equilibrium ensembles, DCC will hardly be produced. On the other hand populating at $\tau_0$ the low momenta fluctuations $n_k$ in a thermal like distribution with a temperature $T_1$ and populating the rest of the higher momenta in a thermal like distribution with a lower temperature $T_2$, we found that instabilities occur even if we don't put kinetic energy in the condensates. Such an initial configuration is close in nature to the quench approximation.

---

[1] We note that for a gaussian density matrix the requirement of isospin invariance leads immediately to a factorization form $\rho = \prod_i \rho_i$. Due to the factorizability of the density matrix, if one species $i$ has a large average $<N^i>$, the whole density matrix will also have a large $<N>$ and vice versa. This is also true for the variance $<(N - \bar{N})^2>$.

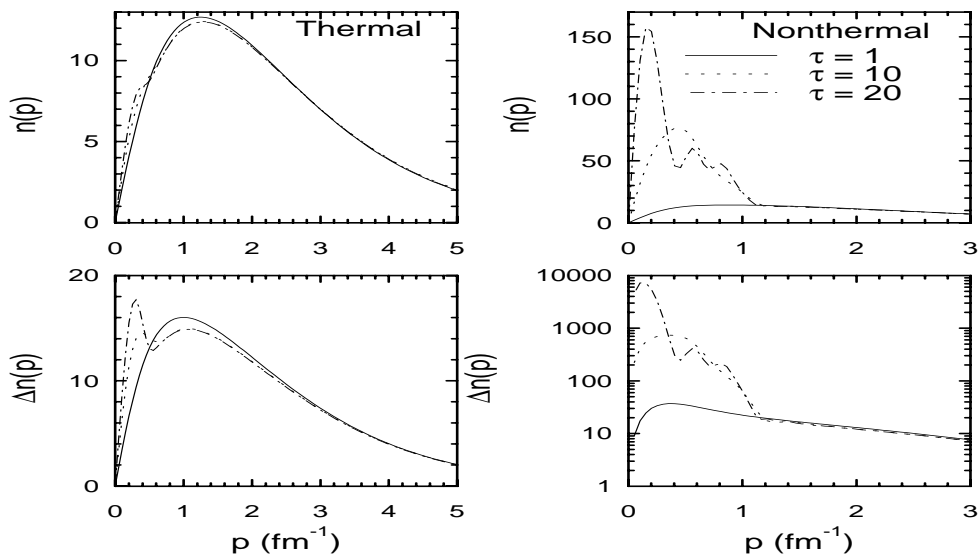

Figure 1. The mean density of particles $<n(p)>$ and the fluctuation $\Delta n(p)$ at $\tau_0 = 1 fm/c$ and at $\tau = 10, 20 fm/c$. The left column is for an initial thermal distribution where $n_k^i = 1/(\exp \beta \omega_k - 1)$, $\bar{\sigma}(\tau_0 = 1) = \bar{\sigma}_{eq}(T = 200 MeV)$, $\bar{\pi}^i(1) = \dot{\bar{\pi}}^i(1) = \dot{\bar{\sigma}}(1) = 0$. The right column has the same initial conditions as in the left one, but for $\dot{\bar{\sigma}}(1) = -1$.

## REFERENCES

1. A. Anselm, Phys. Lett. **B217** (1989) 169; A. Anselm and M. Ryskin, Phys. Lett. **B226** (1991) 482; J.-P. Blaizot, A. Krzywicki, Phys. Rev. **D46** (1992) 246.
2. C.M.G. Lattes, Y. Fujimoto and S. Hasegawa, Phys. Rep. **65** (1980) 151.
3. J.D. Bjorken, K. Kowalsky and C.C. Taylor, SLAC-PUB-6109 preprint (1993).
4. K. Rajagopal and F. Wilczek, Nucl. Phys. **B379** (1993) 395; S. Gavin, A. Gocksch and R.D. Pisarski, Phys. Rev. Lett. **72** (1994) 2143; Z. Huang and X.N. Wang, Phys. Rev. D **49** (1994) R4335; S. Gavin and B. Mueller, Phys. Lett. **B329** (1994) 486.
5. D. Boyanovsky, D.S. Lee and A. Singh, Phys. Rev. **D48** (1993) 800.
6. F. Cooper, Y. Kluger, E. Mottola, J.P. Paz, Phys. Rev. **D51** (1995) 2377.
7. I. Kogan, Phys. ReV. D **48** (1993) 3971; I. Kogan, JETP Lett. **59** (1994) 312.